\documentclass[twocolumn,showpacs,preprintnumbers,amsmath,amssymb]{revtex4}

\usepackage{graphicx}
\usepackage{dcolumn}
\usepackage{bm}

\begin{document}


\title{Paramaterizations of inclusive cross sections for pion production in proton-proton collisions. II. Comparison to new data.}

\author{John W. Norbury}
\email{john. norbury@usm.edu}
\affiliation{Department of Physics and Astronomy, University of Southern Mississippi, Hattiesburg, MS 39406}

\author{Lawrence W. Townsend}
\email{ltownsen@tennessee.edu}
\affiliation{Department of Nuclear Engineering,
University of Tennessee, Knoxville, TN 37996}

\date{\today}

\begin{abstract}

A set of new, precise data have recently been made available by  the NA49 collaboration  for charged  pion production in proton-proton and proton-Carbon reactions at 158 GeV.  The current paper compares this new data to 
 five currently available arithmetic  parameterizations. Although a precise fit is not expected, two of the parameterizations do not work very well but the other three are able to provide a moderately good, but not precise fit to the proton-proton data. The best two of these three parameterizations are scaled to the proton-Carbon data and again provide a moderately good, but not precise fit.

\end{abstract}


\maketitle

\section{INTRODUCTION}

The NA49 collaboration  \cite{altpp, altpc} has recently completed a series of  measurements for charged pion production in proton-proton ($pp$)  and  proton-Carbon ($pC$) collisions at a beam momentum of 158 GeV, corresponding to $\sqrt{s}=17$ GeV. This  surpasses previous data in that the new data is of much higher precision and quality and can therefore be used to provide more precise tests of hadronic production models.  The NA49 collaboration \cite{altpp, altpc} indicate that simple arithmetic parameterizations are unable to account for the fine structure seen in their data, and they therefore provide a numerical interpolation scheme. We agree with this. However arithmetic parameterizations are used in a wide variety of applications  including  simulation of particle physics experiments \cite{fernow}, simulations of cosmic rays showers in the Earth atmosphere \cite{rao, gaisser, sokolsky}, description of particle reactions relevant to astrophysics \cite{longair} and predicting radiation environments inside spacecraft  \cite{wilson}. For those using such parameterizations it is of  interest to know how they compare to the new precision NA49 data  \cite{altpp, altpc}, even though a precision fit will not be possible.  Blattnig, Swaminathan,  Kruger, Ngom, and Norbury  \cite{blattnig} analysed a  set of parameterizations currently available and compared to an extensive data set for both neutral and charged pion production in $pp$ collisions.  It was concluded that the
parameterization of Badhwar and Stephens \cite{badhwar}  provided the best overall description of charged pion production. It is  interesting to see how this compares to the new data \cite{altpp}.

The Blattnig  {\em et al}  parameterizations \cite{blattnig} have also been used in a variety of  astrophysical and astroparticle applications \cite{bernado, santamaria, kelner, kamae-2005, kamae-2006, moskalenko, prodanovic}  where the interest is in calculating the spectrum of gamma rays, electrons and neutrinos which result from the decay of pions  produced in proton-proton interactions. For example, Bernado \cite{bernado} used the high energy $pp \rightarrow \pi^0$ cross sections in order to calculate the $\gamma$-ray spectrum from microquasars. This spectrum can be measured using gamma ray telescopes on  satellites.
Pion production cross section  parameterizations \cite{blattnig} have also been used recently in
nuclear and particle physics applications  \cite{blume, denterria2004, denterria-2005, denterria-a-2005}. 
The nuclear modification factor $R_{AA}$ is basically the ratio of a nucleus-nucleus cross section divided by a scaled nucleon-nucleon cross section. The behavior of $R_{AA}$ can be used to provide information on signatures of quark-gluon plasma formation \cite{blume, denterria2004, denterria-2005, denterria-a-2005}.
Pion parameterizations have been used for the proton-proton cross sections  in these nuclear modification factors \cite{blume, denterria2004, denterria-2005, denterria-a-2005}.
Given such widespread use of pion production cross section  parameterizations, we deem it worthwhile to test currently available  parameterizations against new accelerator data. This is the aim of the present work.

The outline of the paper is as follows. We summarize the paramaterizations studied previously \cite{blattnig} making note of necessary variable transformations needed to describe the  NA49  \cite{altpp, altpc}  data set. We then compare these parameterizations to the new data for $pp$ reactions. Therefore the present work is a continuation of  the previous paper \cite{blattnig} but  applied to the new data.  The parameterizations which are able to give a reasonable fit to the $pp$ data are then scaled and compared to the $pC$ data.

\section{Differential cross sections}

\subsection{Review of Kinematics}

Consider the {\em inclusive} reaction
\begin{eqnarray}
a+b \rightarrow c+X   \label{reaction}
\end{eqnarray}
where $c$ is the produced particle of interest and $X$ is anything. Throughout this paper we assume that all variables, such as all momenta, are evaluated in the  center of momentum (CM) frame.
The momentum of  particle $c$  is denoted as $p$, and supposing  that it comes out at angle $\theta$ to the beam direction, then the longitudinal   and transverse   components of momentum are
\begin{eqnarray}
p_z &\equiv& p \cos \theta \\
p_T &\equiv& p\sin \theta
\end{eqnarray}
Feynman  used a scaled variable instead of $p_z$ itself 
\cite{hagedorn, roe, collins, perkins2}.
The {\em Feynman scaling variable} is   \cite{collins, whitmore, wong, nagamiya, giacomelli, tannenbaum, cassing}
\begin{eqnarray}
x_F  &\equiv& \frac{p_z}{p_{z\; \rm max}}
\end{eqnarray}
where $p_z$ is the longitudinal  momentum of the produced meson in the CM frame, and  $p_{z\; \rm max}$ is the maximum momentum  of the produced meson given by \cite{wong, nagamiya, cassing}
\begin{eqnarray}
p_{z \; \rm max}    &=& \sqrt{\frac{\lambda(s,m_c,m_X)}{4s}}   \label{pmax}
\end{eqnarray}
with
\begin{eqnarray}
\lambda(s,m_i,m_j) \equiv (s-m_i^2-m_j^2)^2 - 4m_i^2 m_j^2
\end{eqnarray}
Note that
\begin{eqnarray}
p_{z \; \rm max}  = p_{\rm max} 
\end{eqnarray}
Nagamiya and Gyulassy \cite{nagamiya} point out that  if $c$ is a boson with zero baryon number, then
 \begin{eqnarray}
 m_X = m_A+m_B 
 \end{eqnarray}
 in agreement with the $p_{z \; \rm max}$  formulas of Nagamiya and Gyulassy \cite{nagamiya} and Cassing \cite{cassing}. The Feynman scaling variable approaches the limiting value \cite{giacomelli}
\begin{eqnarray}
x_F \rightarrow \frac{2p_z}{\sqrt{s}}  \hspace*{2cm} {\rm as} \; s\rightarrow \infty
\end{eqnarray}
Also it is obviously bounded in the following manner \cite{collins}
\begin{eqnarray}
-1 < x_F < 1
\end{eqnarray}
Sets of variables that are often used are either $(p, \theta)$ or $(p_z, p_T)$
Writing
\begin{eqnarray}
p_z =x_F \; \sqrt{\frac{\lambda(s, m_c, m_X)}{4s}}  \label{writing}
\end{eqnarray}
shows that another useful and common variable set is  $(x_F, p_T)$, which  is used by the NA49 collaboration \cite{altpp, altpc} in presenting their data. These variables are also used throughout   the present work.
Rapidity is defined as
\begin{eqnarray}
y = \frac{1}{2} \log \left (  \frac{E+p_z}{E-p_z}\right )
\end{eqnarray}
so that
\begin{eqnarray}
E &=& m_T \cosh y \\
p_z &=& m_T \sinh y
\end{eqnarray}
where the transverse mass is defined through
\begin{eqnarray}
m_T^2 = m^2 + p_T^2 = E^2-p_z^2
\end{eqnarray}
with $m$ as the mass of the produced  particle $c$.This gives yet another useful variable set $(y, p_T)$. In the work below it will be necessary to write the rapidity in terms of the Feynman scaling variable as
\begin{eqnarray}
y = \frac{1}{2} \log \left ( \frac{\sqrt{x_F^2 +m_T^2/p_{z \; \rm max}^2}+x_F}{\sqrt{x_F^2 +m_T^2/p_{z \; \rm max}^2}-x_F}  \right ) \label{yF}
\end{eqnarray}
For massless particles, $E=p$, so that $y$ becomes
\begin{eqnarray}
\eta = \frac{1}{2} \log \frac{1+\cos \theta}{1-\cos \theta} = -\log \left (\tan\frac{\theta}{2} \right )
\end{eqnarray}
This is called the pseudorapidity and is a good approximation to the rapidity for particles moving near the speed of light.  Because the pseudorapdity depends only on angle it can be used as an angular variable. Wong \cite{wong} provides useful formulas involving $\eta$ and also gives expressions relating $y$ to $\eta$ for slower-than-light particles.

\subsection{Parameterizations}

Blattnig {\em et al}  \cite{blattnig} did a  study of the various parameterizations available for inclusive pion production in proton-proton collisions. They concluded that the Badhwar parameterization \cite{badhwar} worked the best for charged pion production.  However other parameterizations  \cite{blattnig, carey, alper, ellis, mokhov} will be reviewed again to see which works best for the new experimental data.  The NA49  data set   \cite{altpp, altpc}  uses the variables $(x_F, p_T)$, whereas some of the parameterizations below are written in terms of other variables sets. These will  need to be converted  to $(x_F, p_T)$.

\subsubsection{Badhwar parameterization }
This parameterization  \cite{badhwar} gives the Lorentz-invariant differential cross section as
\begin{eqnarray}
E\frac{d^3\sigma}{d^3p} &=& \frac{A}{(1+4m_p^2/s)^r} (1-\tilde x)^q \exp[\frac{-Bp_T}{1+4m_p^2/s}]
\end{eqnarray}
where $m_p$ is the proton mass, $\sqrt{s}$ is the total energy in the center of momentum (CM)  frame, and $p_T$ is the transverse momentum of the produced meson in the CM frame. The other terms are given by
\begin{eqnarray}
\tilde x &=& \left [x_F^2 + \frac{4}{s}(p_T^2 +m_\pi^2) \right ]^{1/2} 
\end{eqnarray}
where it is assumed that the variables appearing in $x_F$ are in the CM frame. Badhwar writes 
$x_\parallel^* \equiv x_F$. Also
\begin{eqnarray}
q = \frac{C_1+C_2p_T+C_3p_T^2}{\sqrt{1+4m_p^2/s}}
\end{eqnarray}
The constants are listed in Table 1.  The Badhwar variables are $(x_F, p_T)$, which are the variables used in the NA49 data set  \cite{altpp, altpc}, so that no variable conversion is necessary. \\

\noindent TABLE 1.  Constants for the Badhwar parameterization. Units for $A$,  $C_2$ and $C_3$ are 
$\rm   mb/GeV^2$,  $\rm GeV^{-1}$ and $\rm GeV^{-2}$ respectively,  and other constants are dimensionless.

\noindent
\hrulefill
\begin{tabbing}
xxxxxxxxx\=xxxxxx\=xxxxxxx\=xxxxx\=xxxxxxxxx\=xxxxxx\=xxxxx\kill
Particle    \>\;$A$        \>\;$B$           \>$r$                     \>\;\;$C_1$         \>$C_2$     \>\;$C_3$
\end{tabbing}
\hrulefill
\begin{tabbing}
xxxxxxxxx\=xxxxxx\=xxxxxxx\=xxxxx\=xxxxxxxx\=xxxxxx\=xxxxx\kill
$\;\;\pi^+$      \>153          \>5.55         \>1       \>5.3667         \>-3.5       \>0.8334 \\
 $\;\;\pi^-$      \>127         \>5.3         \>3          \>7.0334         \>-4.5       \>1.667
 \end{tabbing}
\hrulefill  \\

\subsubsection{Alper   parameterization }

The Alper \cite{alper} parameterization used in reference \cite{blattnig} was 
\begin{eqnarray}
E\frac{d^3 \sigma}{d^3p} = A \exp(-B p_T +C p_T^2) \exp(-Dy^2)   \label{alper1}
\end{eqnarray}
where $A$, $B$, $C$, and $D$ are constants that  depend on the value of $\sqrt{s}$.  A more general formula is \cite{alper}
\begin{eqnarray}
E\frac{d^3 \sigma}{d^3p} &=& A_1  \exp(-B p_T) \exp(-Dy^2)   \nonumber \\
&&+A_2 \frac{(1-p_T/p_{\rm beam})^m}{(p_T^2+M^2)^n}   \label{alper2}
\end{eqnarray}
The constants are listed in Table 2.
The Alper variables are $(y, p_T)$.
To change to  the variables $(x_F, p_T)$,  we  convert the rapidity in (\ref{alper2}) to $x_F$ using (\ref{yF}). \\

\noindent TABLE 2.  Constants for  the Alper  parameterization. 

\noindent
\hrulefill
\begin{tabbing}
xxxxxxxx\=xxxxxx\=xxxxxxx\=xxxxx\=xxxxxxx\=xxxxxx\=xxxxx\=xxxxxx\kill
Particle    \>\;$A_1$        \>\;$B$           \>$D$             \>\;\;$A_2$         \>$M$     \>\;$m$  \>$n$
\end{tabbing}
\hrulefill
\begin{tabbing}
xxxxxxxx\=xxxxxx\=xxxxxxx\=xxxxx\=xxxxxxx\=xxxxxx\=xxxxx\=xxxxxx\kill
$\;\;\pi^+$      \>210\>7.58\>0.20\>10.7\>1.03\>10.9\>4.0\\
 $\;\;\pi^-$      \>205\>7.44\>0.21\>12.8\>1.08\>13.1\>4.0
 \end{tabbing}
\hrulefill  \\

\subsubsection{Ellis   parameterization }

The Ellis \cite{ellis} parameterization is 
\begin{eqnarray}
E\frac{d^3 \sigma}{d^3p} =  A(p_T^2 +M^2)^{-N/2}  (1-x_T)^F
\end{eqnarray}
where $A$ is an overall normalization fitted to be $A=13$ in reference \cite{blattnig} and 
$x_T \equiv p_T/p_{\rm max} \approx 2p_T/\sqrt{s}$.  The same value of $A$ is used in the present work. The other constants are listed in Table 3.
The Ellis parameterization is independent of the emission angle $\theta$, and so does not carry any dependence on $p_z$, $x_F$, $y$ etc. \\

\noindent TABLE 3.  Constants for  the Ellis  parameterization. 

\noindent
\hrulefill
\begin{tabbing}
xxxxxxxxxxxxxxx\=xxxxxxxxxxxxx\=xxxxxxxxxxxxxx\=xxxxxxxxxxx\kill
Particle    \>\;$N$ \>\;$M^2$ \>\;$F$
\end{tabbing}
\hrulefill
\begin{tabbing}
xxxxxxxxxxxxxxx\=xxxxxxxxxxxxx\=xxxxxxxxxxxxxx\=xxxxxxxxxxx\kill
$\;\;\pi^+$      \>7.70\>0.74\>11.0  \\
 $\;\;\pi^-$      \>7.78\>0.79\>11.9
 \end{tabbing}
\hrulefill  \\

\subsubsection{Mokhov  parameterization }
The Mokhov  \cite{mokhov} parameterization is 
\begin{eqnarray}
E\frac{d^3 \sigma}{d^3p} =  A \left( 1-\frac{p}{p_{\rm max}} \right )^B \exp \left (-\frac{p}{C\sqrt{s}} \right )
V_1(p_T)V_2(p_T)  \nonumber
\end{eqnarray}
where
\begin{eqnarray}
V_1(p_T) &=&  (1-D) \exp(-Ep_T^2)+D \exp(-Fp_T^2) \nonumber \\
&&\hspace*{2cm}\;\;{\rm for} \;p_T \leq 0.933 \; {\rm GeV}   \nonumber \\
&=&  \frac{0.2625}{(p_T^2+0.87)^4}
\;\;{\rm for} \;p_T  > 0.933 \; {\rm GeV}  \nonumber 
\end{eqnarray}
and
\begin{eqnarray}
V_2(p_T) &=&  0.7363 \exp(0.875 p_T)   \;\;{\rm for} \;p_T \leq 0.35 \; {\rm GeV}  \nonumber  \\
 &=& 1   \;\;{\rm for} \;p_T > 0.35 \; {\rm GeV}  \nonumber 
\end{eqnarray}
The constants are listed in Table 4. 
Using  $p=\sqrt{p_z^2+p_T^2}$,  gives  the Mokhov variables $(p_z, p_T)$ which are transformed to $(x_F, p_T)$ using (\ref{writing}). \\

\begin{figure}
\begin{center}
\includegraphics[width=3in]{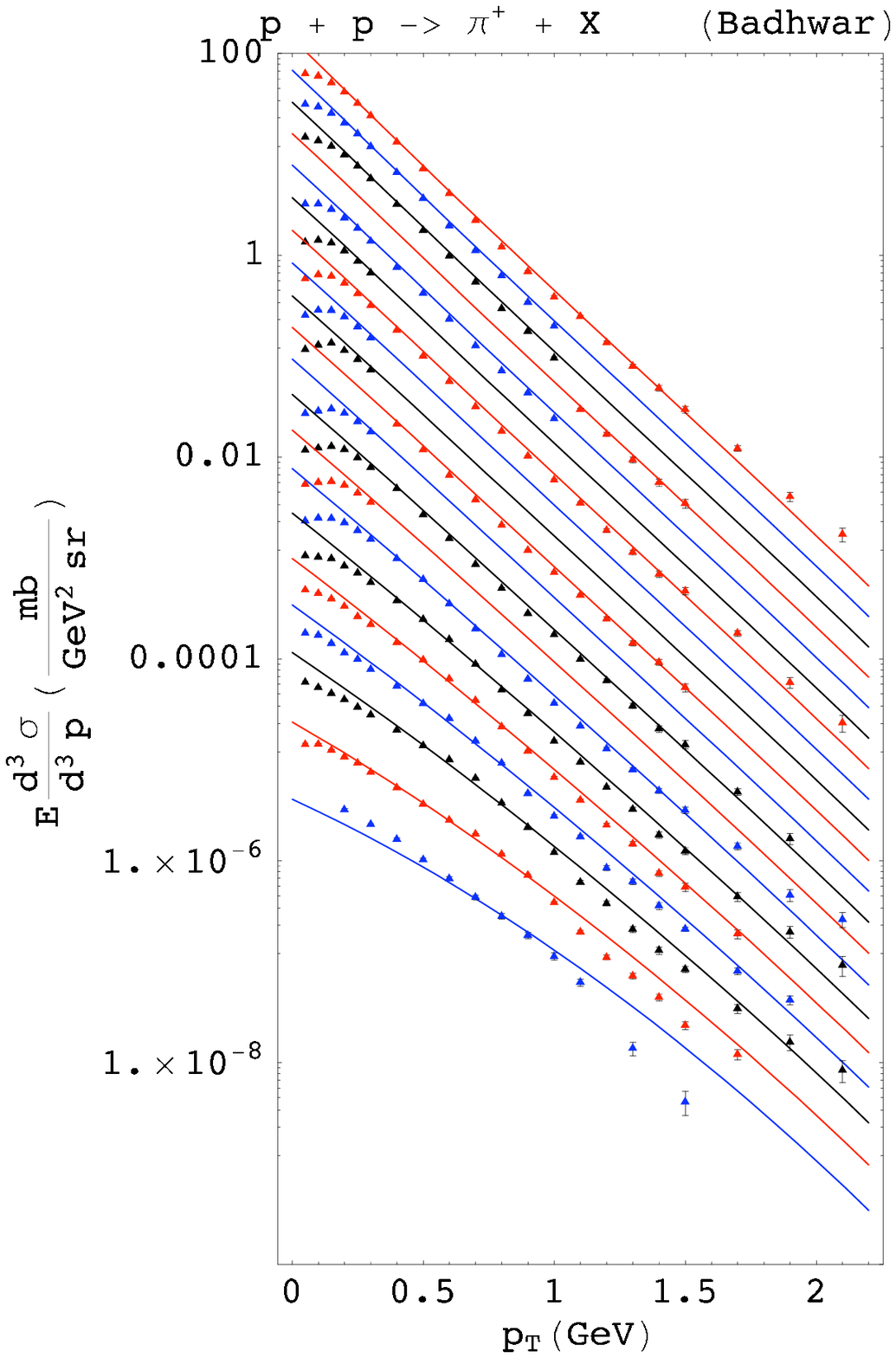} 
 \end{center}
\caption{ $\pi^+$ production in $pp$ collisions. Data from reference \cite{altpp} is plotted against the parameterization of Badhwar \cite{badhwar}.  The values of $x_F$ from top to bottom are 0.0,  0.01,  0.02,  0.025,  0.03,  0.04,  0.05,  0.06,  0.07,  0.075,  0.08,  0.1,  0.12,  0.15,  0.2,  0.25,  0.3,  0.35,  0.45,  0.55. Following \cite{altpp}, data and lines are multiplied successively by 0.5 to allow for a better separation. }
\end{figure}

\noindent TABLE 4.  Constants for  the Mokhov parameterization. 

\noindent
\hrulefill
\begin{tabbing}
xxxxxxxxxx\=xxxxxx\=xxxxxxx\=xxxxxxx\=xxxxxxx\=xxxxxx\=xxxxxxxx\kill
Particle    \>\;$A$        \>\;$B$           \>\;$C$             \>\;$D$         \>$E$     \>\;$F$ 
\end{tabbing}
\hrulefill
\begin{tabbing}
xxxxxxxxxx\=xxxxxx\=xxxxxxx\=xxxxxxx\=xxxxxxx\=xxxxxx\=xxxxxxxx\kill
$\;\;\pi^+$      \>60.1\>1.9\>0.18\>0.3\>12\>2.7 \\
 $\;\;\pi^-$     \>51.2\>2.6\>0.17 \>0.3\>12\>2.7 
 \end{tabbing}
\hrulefill  \\

\subsubsection{Carey  parameterization }
The Carey \cite{carey} parameterization, which only applies to $\pi^-$,  is
\begin{eqnarray}
E\frac{d^3 \sigma}{d^3p}  (\pi^-) =  N(p_T^2 +0.86)^{-4.5} (1-x_R)^4
\end{eqnarray}
where $N$ is an overall normalization fitted to be $N=13$ in reference \cite{blattnig} and 
$x_R \equiv p/p_{\rm max} \approx 2p/\sqrt{s}$.   The same value of $N$ is used in the present work.  The Carey  variables are $(p_z, p_T)$.
To change to  the variables $(x_F, p_T)$, we use
$x_R = \sqrt{x_F^2 + p_T^2/p_{\rm max}^2}$.

\section{COMPARISON TO DATA}

The above parameterizations are compared to the new experimental data in Figs. 1 - 13. 

\subsection{Comparison to $pp$ data}

 Positive pion production in $pp$ reactions  is shown in Figs. 1 - 4.
In agreement with the conclusions of  the NA49 collaboration  \cite{altpp}, none of these  arithmetic parameterizations is able to account for all the fine structure seen in the data.  The Badhwar, Alper and Ellis parameterizations  are unable to reproduce the shape of the differential cross section at low $p_T$.
The Alper and Ellis parameterizations are unable to account for the data at larger values of $x_F$. However, on the positive side, the Mokhov parameterization (Fig. 4) does give a reasonable description of the shape at low $p_T$ and the Badhwar parameterization (Fig. 1) gives a reasonable  fit to the data over all $x_F$ values for $p_T > 0.3$ GeV. In general, the Alper and Ellis fits are not good, but the Badhwar and Mokhov fits are moderately good, but certainly not precise.

\begin{figure}
\begin{center}
\includegraphics[width=3in]{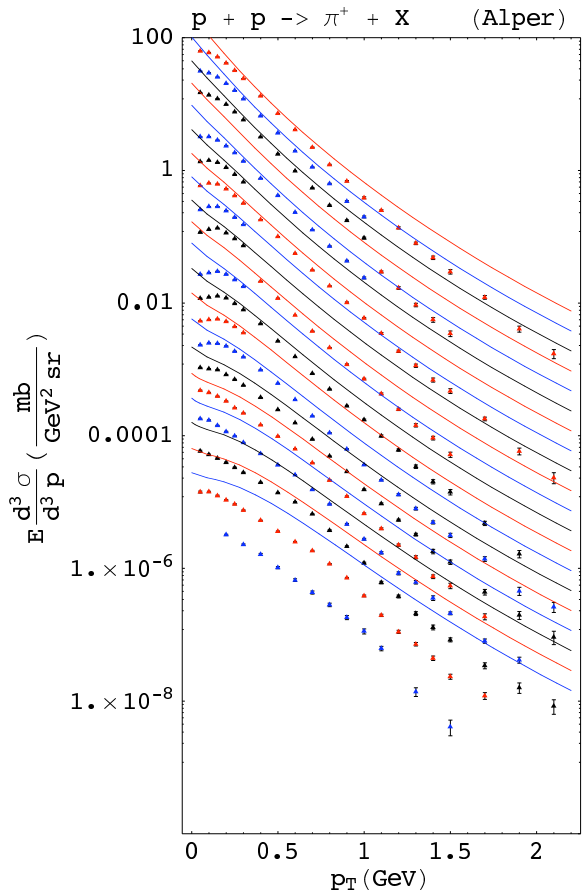} 
 \end{center}
\caption{$\pi^+$ production in $pp$ collisions. Data from reference \cite{altpp} is plotted against the parameterization of Alper  \cite{alper}. Values of $x_F$ and data and line multiplication are the same as Fig.1.}
\end{figure}

\begin{figure}
\begin{center}
\includegraphics[width=3in]{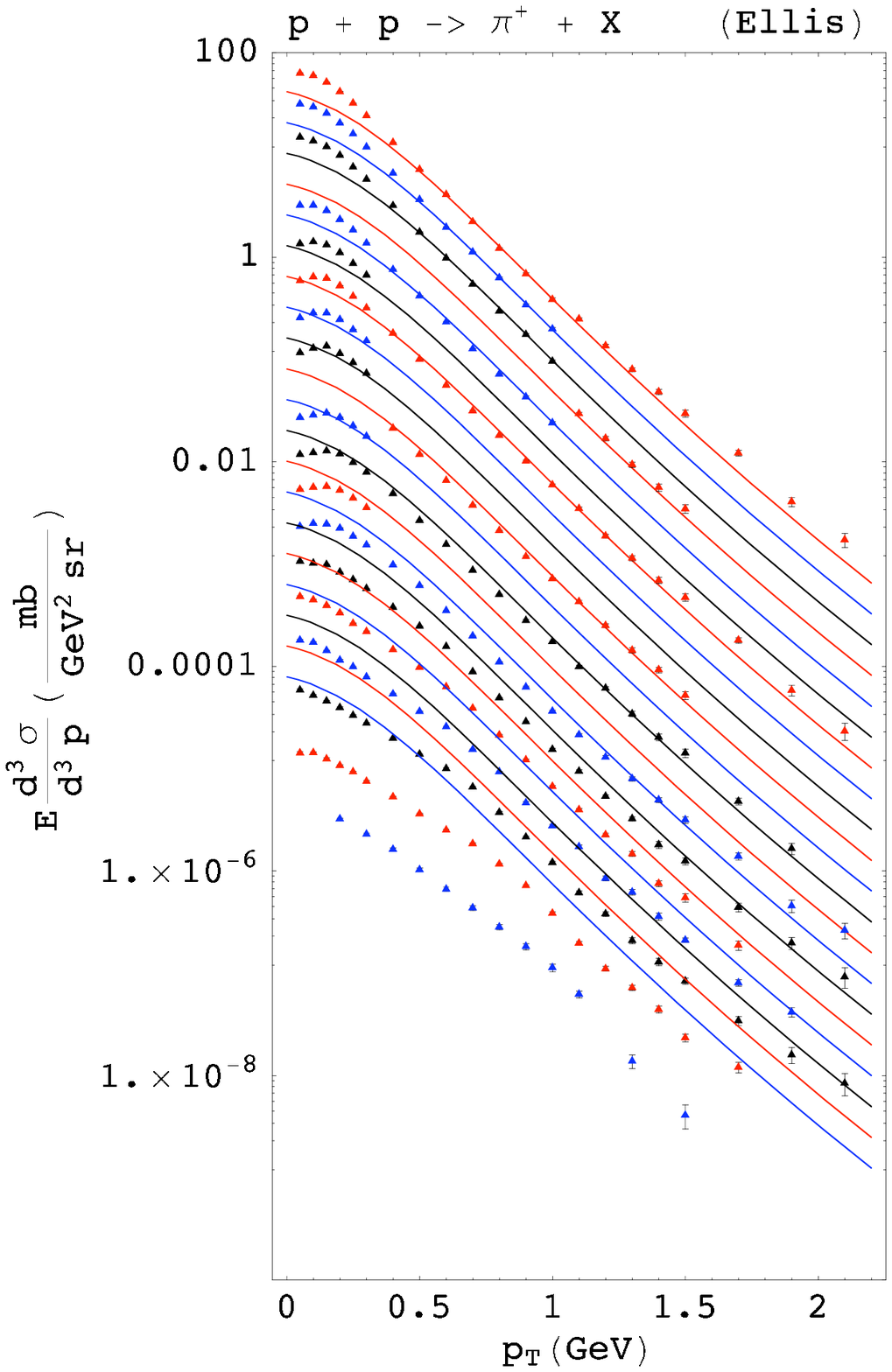} 
 \end{center}
\caption{$\pi^+$ production in $pp$ collisions. Data from reference \cite{altpp} is plotted against the parameterization of Ellis \cite{ellis}. Values of $x_F$ and data and line multiplication are the same as Fig.1.}
\end{figure}

\begin{figure}
\begin{center}
\includegraphics[width=3in]{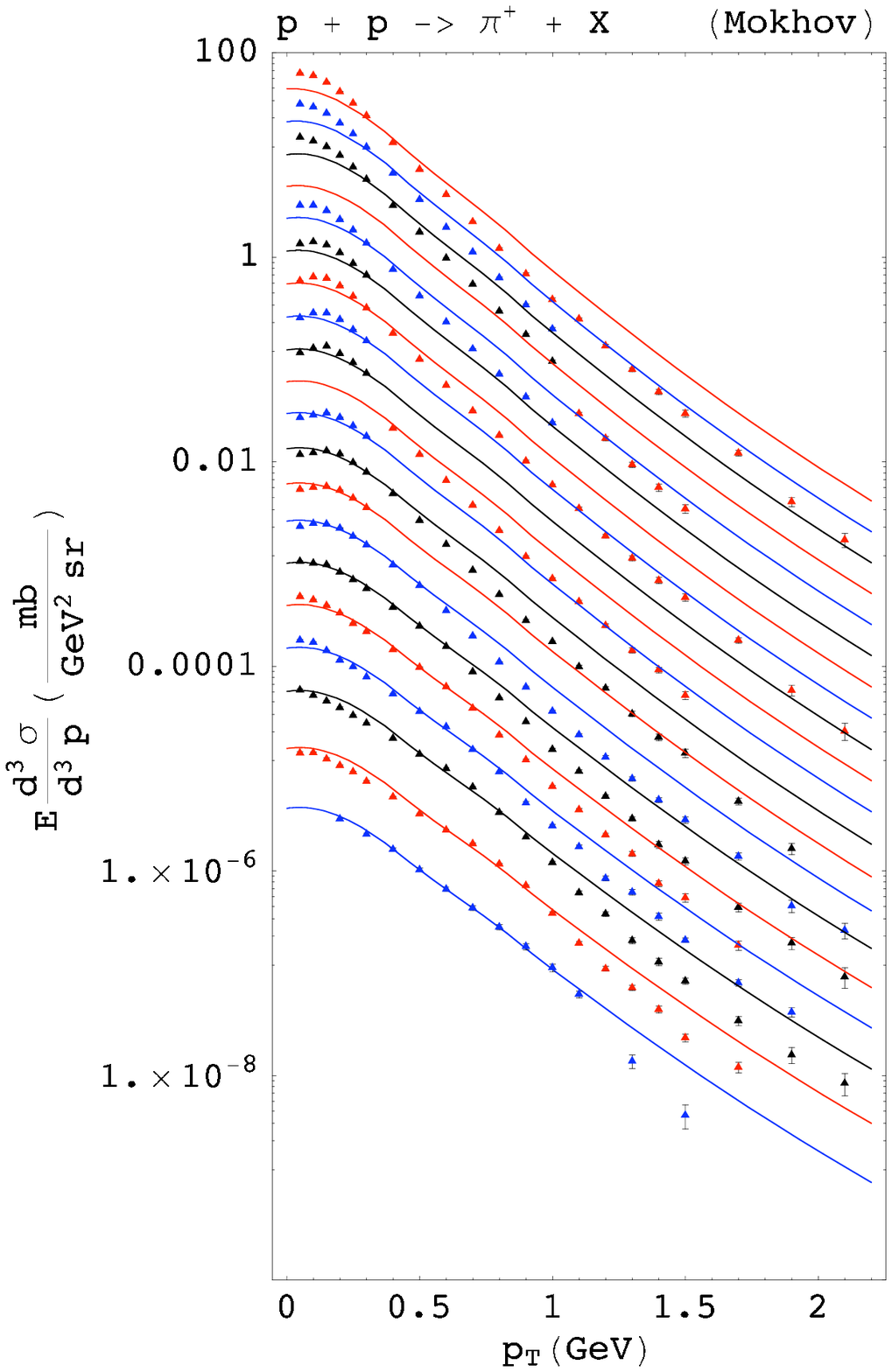} 
 \end{center}
\caption{$\pi^+$ production in $pp$ collisions. Data from reference \cite{altpp} is plotted against the parameterization of Mokhov \cite{mokhov}. Values of $x_F$ and data and line multiplication are the same as Fig. 1.}
\end{figure}

\begin{figure}
\begin{center}
\includegraphics[width=3in]{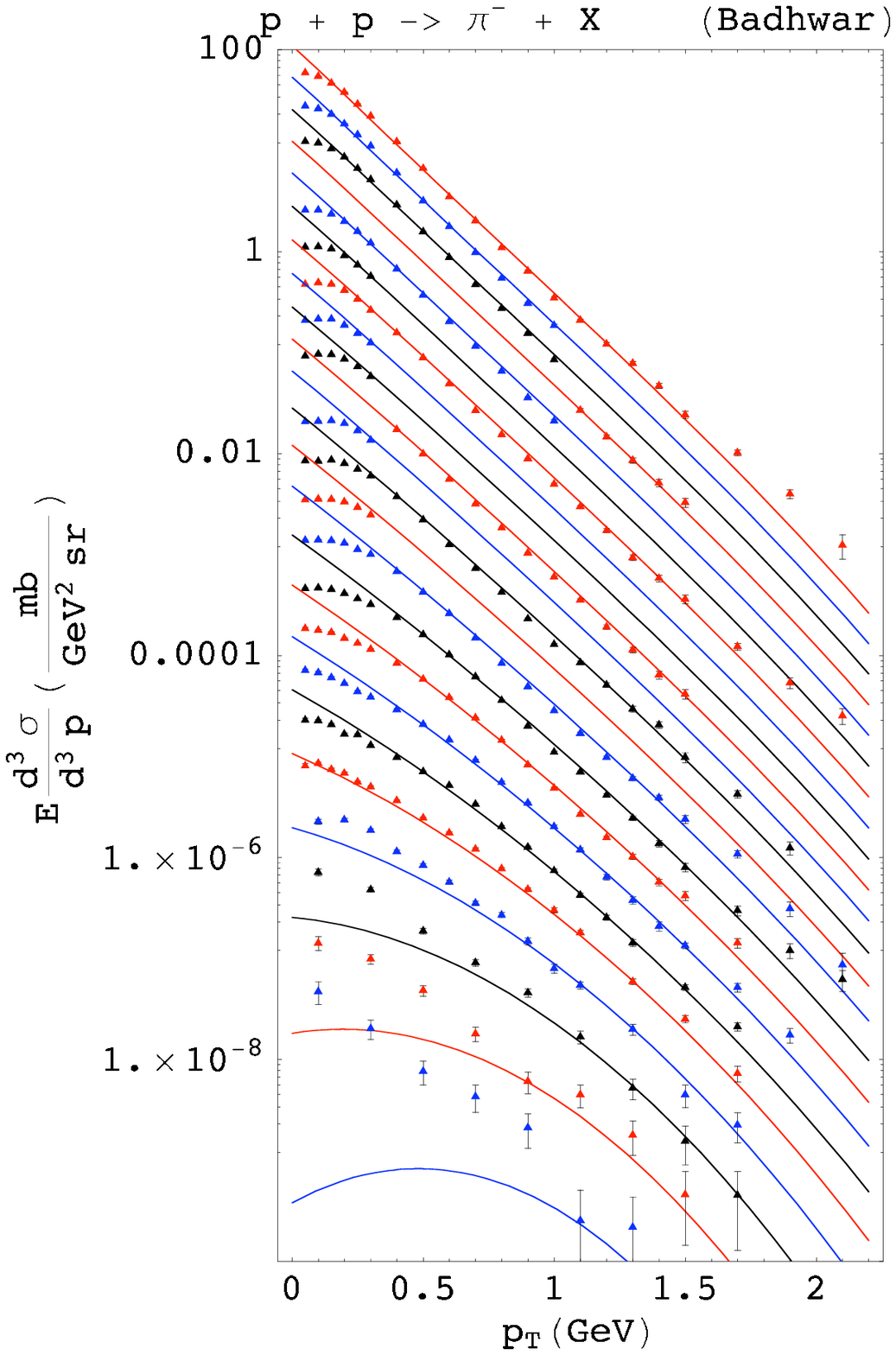} 
 \end{center}
\caption{$\pi^-$ production in $pp$ collisions. Data from reference \cite{altpp} is plotted against the parameterization of  Badhwar \cite{badhwar}.   The values of $x_F$ from top to bottom are  0.0,  0.01,  0.02,  0.025,  0.03,  0.04,  0.05,  0.06,  0.07,  0.075,  0.08,  0.1,  0.12,  0.15,  0.2,  0.25,  0.3,  0.35,  0.45,  0.55, 0.65, 0.75, 0.85. Following \cite{altpp}, data and lines are multiplied successively by 0.5 up to $x_F=0.35$ and by 0.75 for $x_F \geq 0.45$ to allow for a better separation. }
\end{figure}

Negative  pion production in $pp$ reactions  is shown in Figs. 5 - 9.
Again, in  agreement with the conclusion of reference \cite{altpp}, none of these  arithmetic parameterizations is able to account for all the fine structure seen in the data.  The conclusions are the same as for positive pions production, except that the Badhwar  parameterization now fails for larger  values of $x_F$. The Carey  parameterization is reasonable except for  small values of $x_F$ at low $p_T$. In general, the Alper and Ellis fits are not good, but the Badhwar, Mokhov and Carey  fits are moderately good, but not precise.

\begin{figure}
\begin{center}
\includegraphics[width=3in]{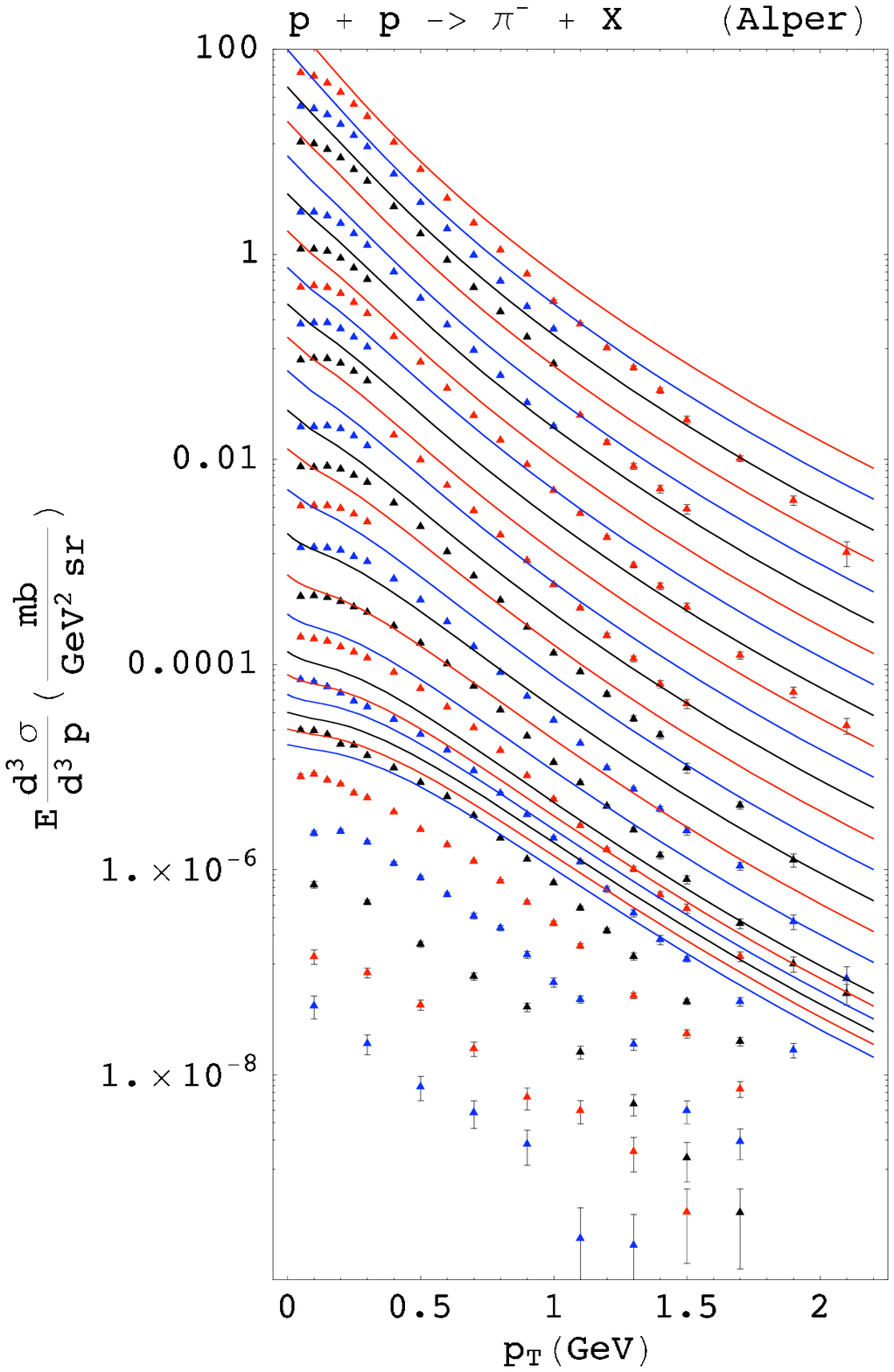} 
 \end{center}
\caption{$\pi^-$ production in $pp$ collisions. Data from reference \cite{altpp} is plotted against the parameterization of  Alper  \cite{alper}. Values of $x_F$ and data and line multiplication are the same as Fig.5.}
\end{figure}

\begin{figure}
\begin{center}
\includegraphics[width=3in]{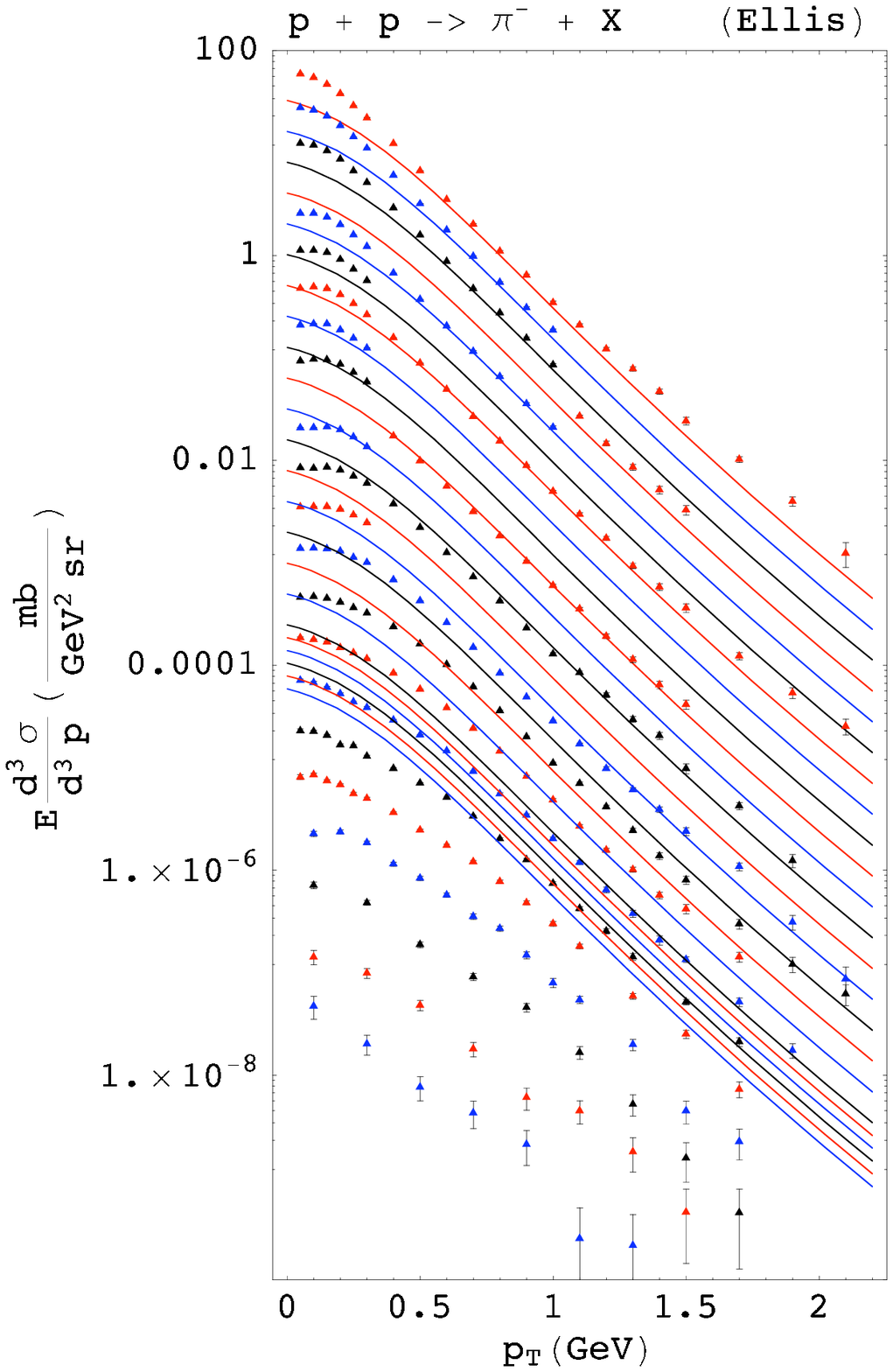} 
 \end{center}
\caption{ $\pi^-$ production in $pp$ collisions. Data from reference \cite{altpp} is plotted against the parameterization of  Ellis  \cite{ellis}.  Values of $x_F$ and data and line multiplication are the same as Fig.5.}
\end{figure}

\begin{figure}
\begin{center}
\includegraphics[width=3in]{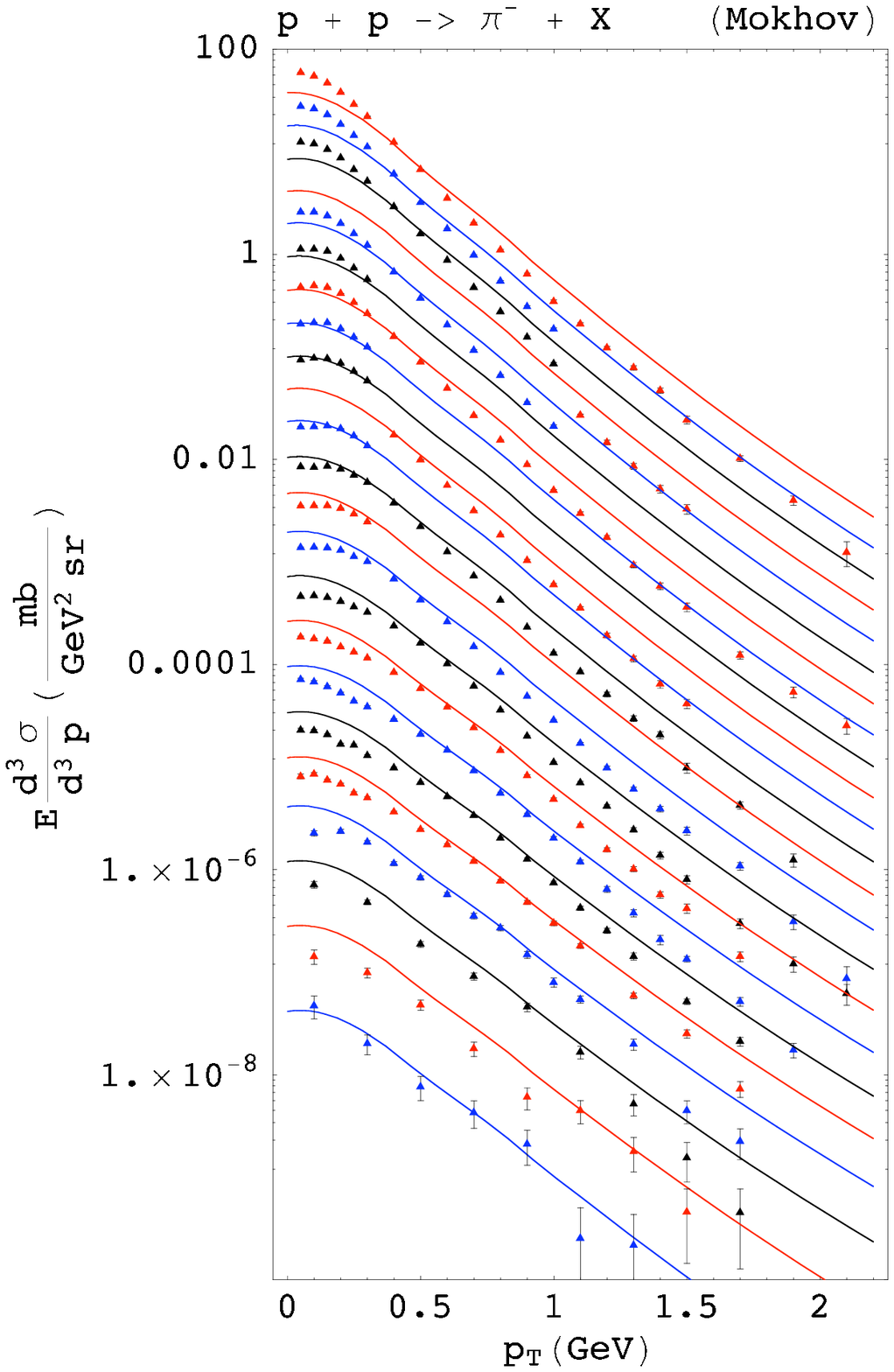} 
 \end{center}
\caption{ $\pi^-$ production in $pp$ collisions. Data from reference \cite{altpp} is plotted against the parameterization of  Mohov  \cite{mokhov}.  Values of $x_F$ and data and line multiplication are the same as Fig.5.}
\end{figure}

\begin{figure}
\begin{center}
\includegraphics[width=3in]{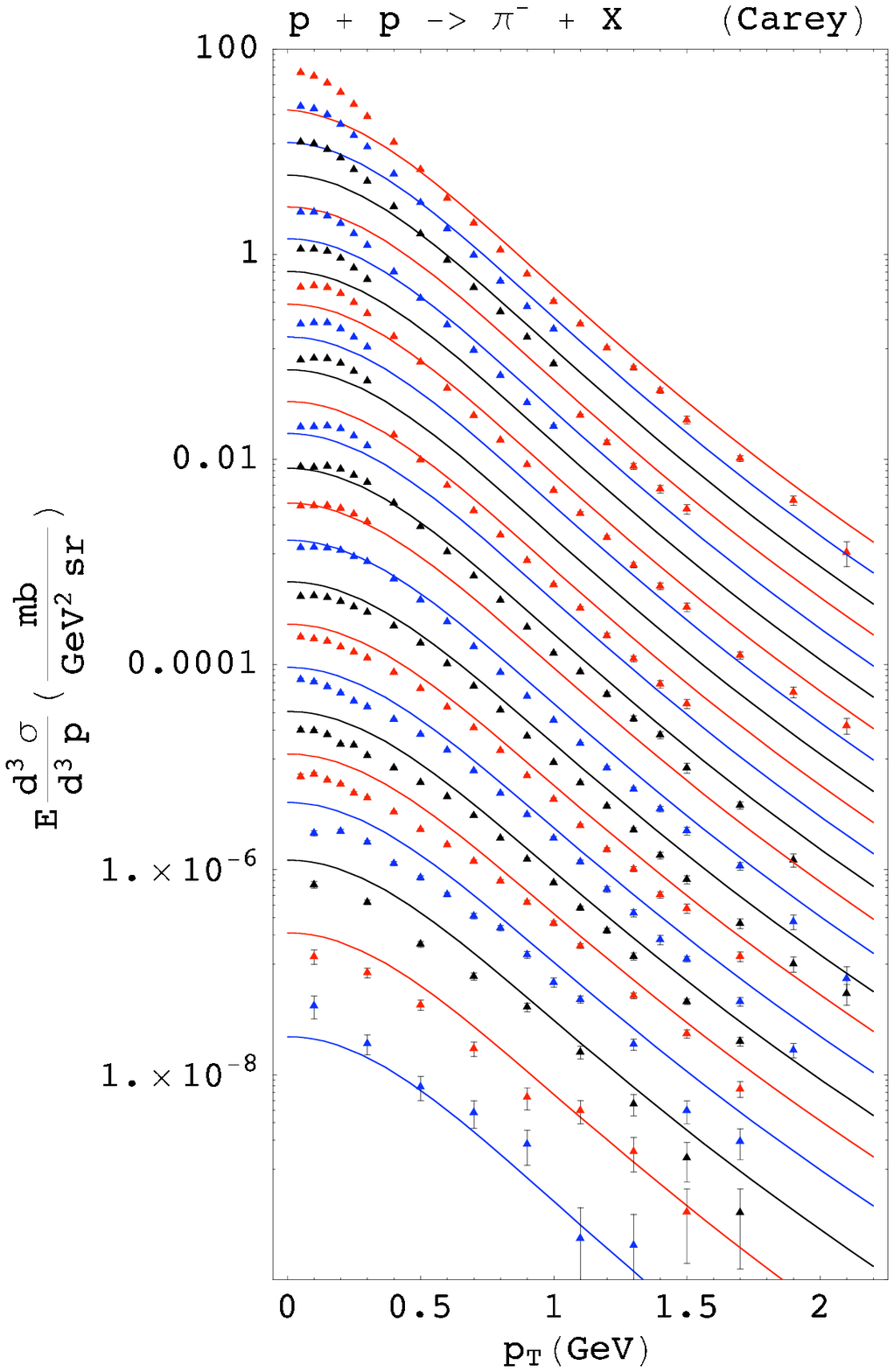} 
 \end{center}
\caption{ $\pi^-$ production in $pp$ collisions. Data from reference \cite{altpp} is plotted against the parameterization of  Carey  \cite{carey}.  Values of $x_F$ and data and line multiplication are the same as Fig.5.}
\end{figure}

\begin{figure}
\begin{center}
\includegraphics[width=3in]{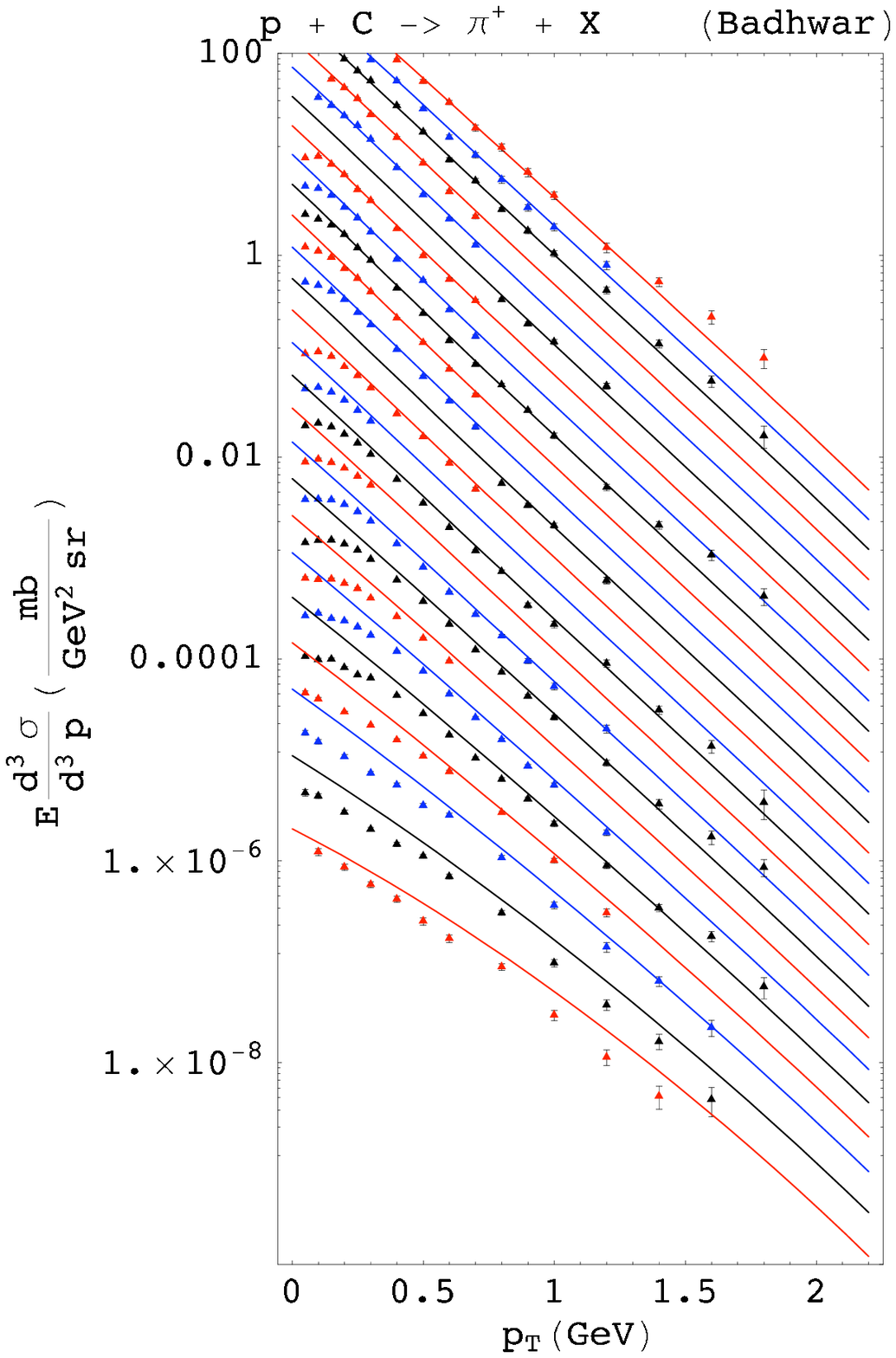} 
 \end{center}
\caption{ $\pi^+$ production in $pC$ collisions. Data from reference \cite{altpc} is plotted against the parameterization of  Badhwar \cite{badhwar} multiplied by a best fit  factor of   
$12^{0.9}$.
The values of $x_F$ from top to bottom are -0.1, -0.075, -0.05,  -0.04, -0.03, -0.025, -0.02, -0.01, 0.0, 0.01, 0.02, 0.025, 0.03, 0.04, 0.05, 0.06, 0.075, 0.1, 0.125, 0.15, 0.2, 0.25, 0.3, 0.4, 0.5. Following \cite{altpp}, data and lines are multiplied successively by 0.5 to allow for a better separation. }
\end{figure}

\subsection{Comparison to $pC$ data}
Although the primary emphasis of this paper is on the $pp$ reactions, nevertheless it is of interest to see if the $pp$ parameterizations are able to scale to fit the new  $pC$ data \cite{altpc}, although a precise fit is not expected.  We will only  consider the parameterizations of  Badhwar and Mokhov  because they provided the best fits to the  $pp$ data.

\begin{figure}
\begin{center}
\includegraphics[width=3in]{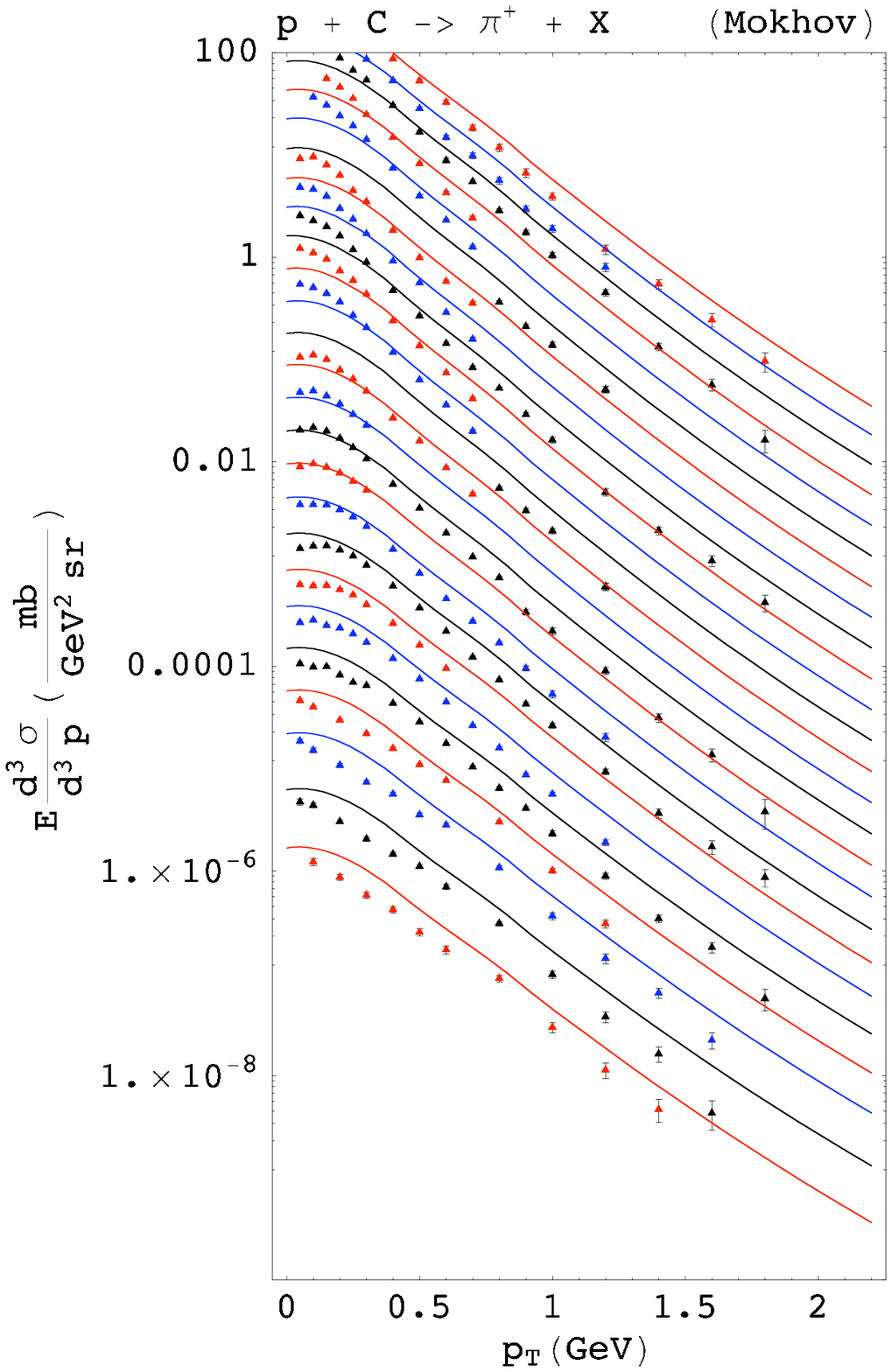} 
 \end{center}
\caption{$\pi^+$ production in $pC$ collisions. Data from reference \cite{altpc} is plotted against the parameterization of  Mokhov \cite{mokhov} multplied by a best fit factor of   
$12^{0.9}$.
Values of $x_F$ and data and line multiplication are the same as Fig. 10.}
\end{figure}

One should note that in comparing to the $pC$ data we will assume
the same cross
 ÊÊ section for proton-proton and proton-neutron  scattering. However, as shown in the paper by Pawlowski and Szczurek  \cite{pawlowski}, this is 
  known not to be true in general for 
 ÊÊ the energy region of the new NA49 data. We justify our use of the same cross sections by noting the following. The data of Pawlowski and Szczurek  \cite{pawlowski} (see their Fig. 2), show that these cross sections are significantly different only for $x_F < -0.05$. For $x_F >-0.05$ the cross sections are very similar and in fact for $x_F >0$,  Pawlowski and Szczurek  \cite{pawlowski} themselves state that  $\sigma_{pp}^{\pi^\pm} \approx  \sigma_{pn}^{\pi^\pm}$.  In our comparisons, we show 25 different values of $x_F$. Only two of these, namely $x_F = -0.01$ and $x_F = -0.075$ are such that there is a significant difference between the proton-proton and proton-neutron cross section. For $\pi^+$ production the proton-proton cross section is about twice that of the proton-neutron cross section and vice-versa for $\pi^-$ \cite{pawlowski}. This means that  only for $x_F = -0.01$ and $x_F = -0.075$ our comparisons to data will be slightly worse for $\pi^+$ and slightly better for $\pi^-$.
We now discuss various models.

\begin{figure}
\begin{center}
\includegraphics[width=3in]{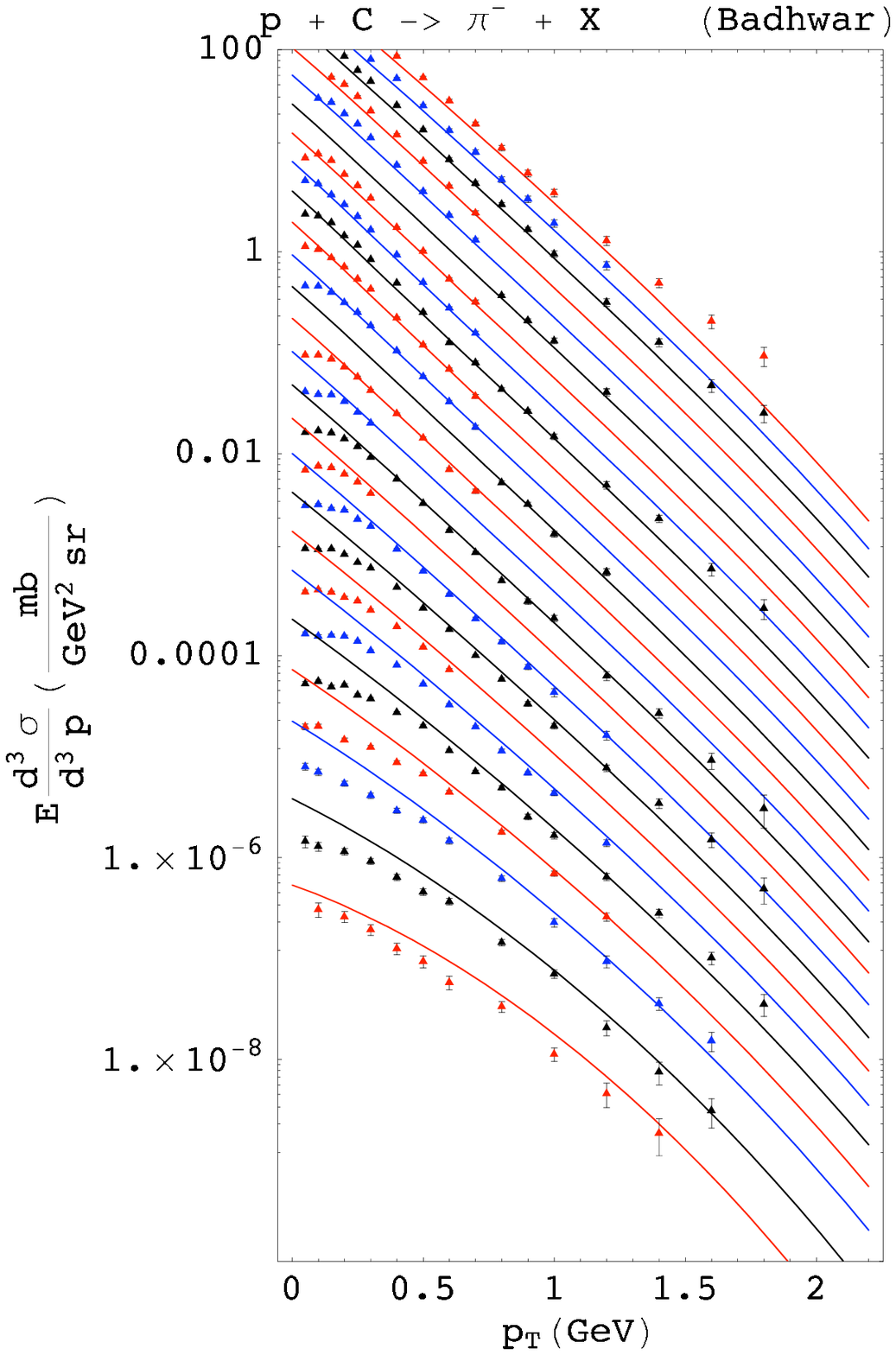} 
 \end{center}
\caption{  $\pi^-$ production in $pC$ collisions. Data from reference \cite{altpc} is plotted against the parameterization of  Badhwar  \cite{badhwar} multplied by a best fit  factor of   
$12^{0.9}$. Values of $x_F$ and data and line multiplication are the same as Fig. 10.}
\end{figure}

\begin{figure}
\begin{center}
\includegraphics[width=3in]{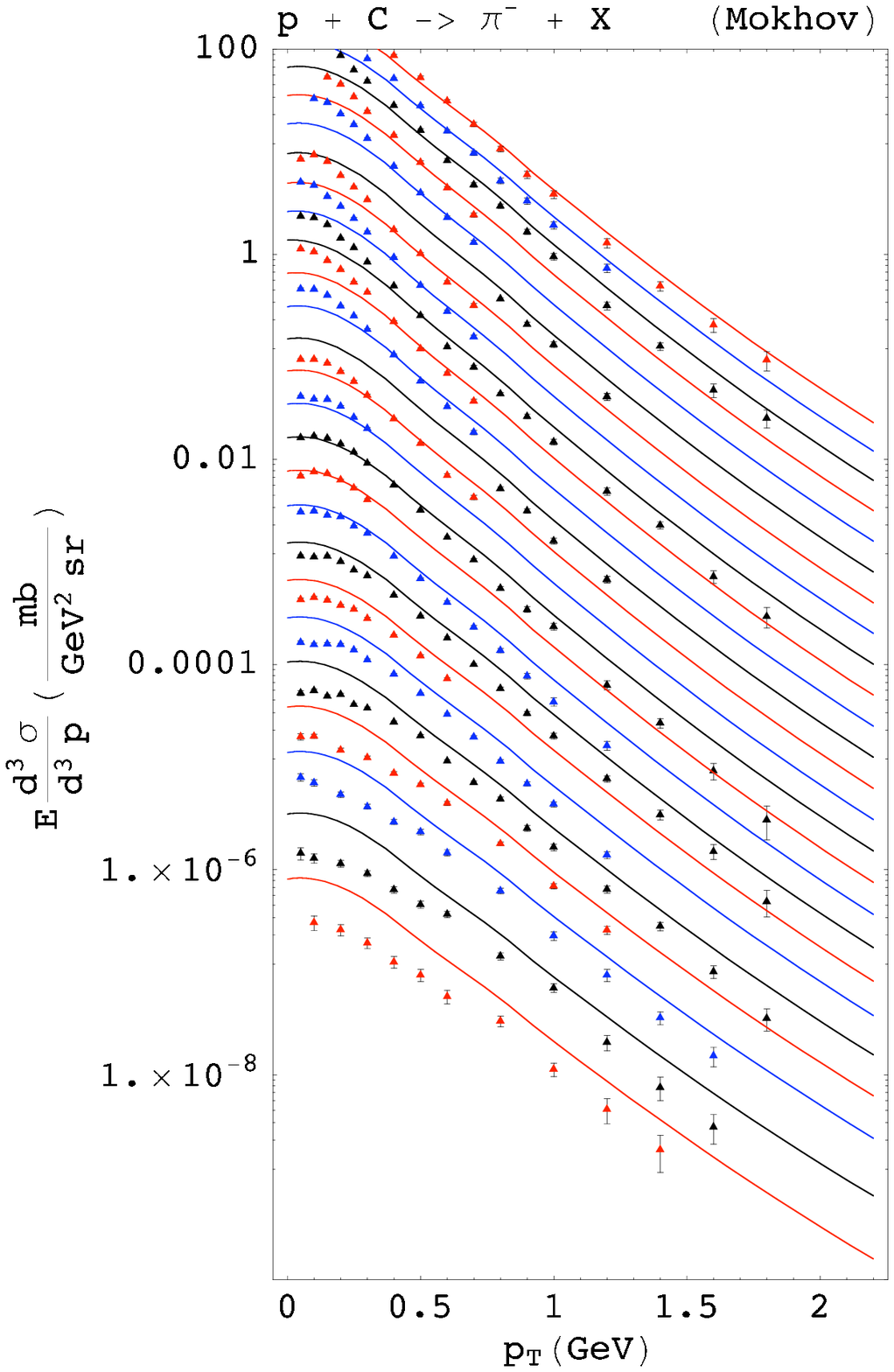} 
 \end{center}
\caption{$\pi^-$ production in $pC$ collisions. Data from reference \cite{altpc} is plotted against the parameterization of  Mokhov  \cite{mokhov} multiplied by a best fit  factor of   
$12^{0.9}$. Values of $x_F$ and data and line multiplication are the same as Fig. 10.}
\end{figure}

 The {\em wounded nucleon model} was introduced by Bialas, Bleszynski and Czyz \cite{bialas}. 
 The number of wounded nucleons is simply the number of participants involved in the reaction.  The main assumption  is that the particle multiplicty is proportional to the number of wounded nucleons.
 Consider $pA$ reactions.  The incident proton interacts with $\nu$ nucleons in the target nucleus \cite{soltz}, which is  some fraction of the total nucleon number $A$, and is determined from  collision geometry  and the hadron-nucleon cross section \cite{soltz}. 
  In a $pA$ reaction, the  number of participants is $(1+\nu)$. Let $N_{pA}$ be the multiplicity of particles of interest  produced in the $pA$ reaction, and let  $N_{pp}$ be the particle multiplicity in the   $pp$ reaction. For example this could be the number of pions produced in a reaction. The multiplicities are related by
$N_{pA} = \frac{1}{2} (1+\nu) N_{pp}$
The $\frac{1}{2}$ factor is there because the ``nucleon-nucleon interaction requires two wounded nucleons"  \cite{navia}.
  The following comes from reference \cite{navia}.
  For a nucleus-nucleus collision let  $A$ be the number of nucleons in the projectile and let $B$ be the number of nucleons in the target. Then \cite{navia} 
$ N_{AB} = \frac{1}{2} W_{AB} N_{pp} $
  where $W_{AB}$ is the number of participant nucleons, i.e. the ``number of nucleons that have interacted at least once" \cite{navia}. This is given by \cite{navia}
$W_{AB} = A \frac{\sigma_{pB}}{\sigma_{AB}} + B \frac{\sigma_{pA}}{\sigma_{AB}}$
  where $\sigma_{pA}$ and   $\sigma_{pB}$ are the proton-nucleus inelastic cross sections and $\sigma_{AB}$ is the nucleus-nucleus cross section given in reference \cite{navia}. If the projectile nucleus is actually a proton then
$W_{pB} =   \frac{\sigma_{pB}}{\sigma_{pB}} + B \frac{\sigma_{pp}}{\sigma_{pB}} 
= 1+ B \frac{\sigma_{pp}}{\sigma_{pB}}$
  or
$1+\nu = W_{pA} =1+ A \frac{\sigma_{pp}}{\sigma_{pA}}$
  showing that \cite{bialkowska}
$\nu =  A \frac{\sigma_{pp}}{\sigma_{pA}}$.
 In $pA$ or $AA$ collisions then the particle multiplicity scales with the number of wounded nucleons, which is often  calculated with Glauber theory \cite{huang}.

 Instead of wounded nucleon or participant  scaling, one can have {\em binary scaling}.
The total number of produced particles  in  nuclear  collisions comes from both soft and hard processes. Soft processes should scale with the number of participants  and hard processes should scale \cite{yin} with  the average number of binary nucleon-nucleon collisions denoted as $N_{\rm coll}$.
 Reference \cite{adcox} contains a nice introduction to {\em hard processes}.  For example, the PHENIX experiment \cite{adcox} focuses on detecting large transverse momentum $p_T$ particles that arise from the early stages of relativistic heavy ion collisions. In the early stage nucleon-nucleon collisions cause jet production resulting from hard parton collisions. The jets subsequently decay into high momentum hadrons with $p_T \sim 2$ Gev \cite{adcox}. The quark-gluon plasma (QGP) forms at a later time in the collision and the early scattered partons move through the  QGP region leading to the phenomenon known as {\em jet quenching} which is signified by a ``depletion in the yield of high $p_T$ hadrons" \cite{adcox}. Observation of this jet quenching is therefore a ``potential signature for QGP formation."

The {\em nuclear modification factor}  is a measure of  nuclear effects and has been discussed by many authors \cite{blume, szczurek, yin, adams, adler, adcox}.
For proton-nucleus reactions, 
 \begin{eqnarray}
 R_{pA} = \frac{Ed^3\sigma_{pA}/d^3p}{N_{\rm coll}\;Ed^3\sigma_{pp}/d^3p}
 \end{eqnarray}
For nucleus-nucleus reactions, 
  \begin{eqnarray}
 R_{AB} = \frac{Ed^3\sigma_{AB}/d^3p}{N_{\rm coll}  \;Ed^3\sigma_{NN}/d^3p}
 \end{eqnarray}
 where  $NN$ refers to the nucleon-nucleon cross section. Instead of $d^3p$, other variables can be used such as $x_F$, $p_T$, rapidity, psuedorapidity, etc.
  The nuclear modification factor $R = 1$ in the absence of nuclear effects, i.e.  if the nuclear cross section is  just an  incoherent superposition of nucleon-nucleon collisions.
  For low $p_T < 2$ GeV it has been found \cite{adcox} that  $R <1$ and this is due to the fact that the reactions scales with the number of participants (participant scaling), rather than the number of binary collisions \cite{adcox}. For $p_T > 2$ GeV   ``particle production in $pA$ collisions is enhanced compared to binary scaling" \cite{adcox}, which is the Cronin effect.

Again  consider {\em participant scaling versus binary scaling.}  The nucleus-nucleus cross section could scale either as a function of the number of participants (often called wounded nucleons) or as a function of the number of binary nucleon-nucleon collisions. Binary scaling would indicate no collective effects whatsoever. However if there are some collective effects then these should manifest themselves by scaling with the number of participants. Obviously binary collisions are harder and lead to high $p_T$ processes such as 
 jet production (from individual parton collisions) and heavy flavor production. Participant effects are generally softer processes at smaller $p_T$ such as soft hadron production, transverse energy flow and other collective phenomena.
 Both the number of participants $N_{\rm part}$ and the number of binary collisions $N_{\rm coll}$ is large for small impact parameter (central collisions) and drops off smoothly at larger impact parameters (peripheral collisions).  Also we always have  $N_{\rm coll} > N_{\rm part}$, because there can be many re-scatterings among the participants.

For a long time it has been known that energetic particle production in proton-nucleus collisions increases faster than the number of  binary nucleon-nucleon collisions \cite{adler}. In other words, 
particle production in nuclear collisions is enhanced compared to binary scaling \cite{adcox}.  This is known as the {\em Cronin effect}. Various physical mechanisms can contribute to the Cronin effect such as multiple parton scattering in the initial stage of the collision \cite{yin}, Fermi motion \cite{szczurek}, etc. Reference \cite{adler} points out that the ``the cause of the Cronin effect and its species dependence are not yet completely understood", where species dependance refers to the fact that the size of the  effect varies for different particles produced. A contribution from final state interactions is also possible \cite{adler}. Reference \cite{adler} provides a very nice summary of the Cronin effect. The $pA$ cross section is parameterized as \cite{adler, vogt}
\begin{eqnarray}
E\frac{d^3\sigma }{d^3p} (p_T,A) = E\frac{d^3\sigma }{d^3p} (p_T,1) \;A^{\alpha(p_T)}
\end{eqnarray}
 where $ \alpha > 1$ indicates nuclear collective effects.
 Adler et al \cite{adler} state that ``the enhancement depends on the momentum and the type of particle produced, with protons and antiprotons exhibiting a much larger enhancment than pions and kaons at $p_T > 2-3$ GeV." They also point out that at $\sqrt{s}=27.4$ GeV, the enhancement reaches a maximum at $p_T=4.5$ GeV with $\alpha_K^+ \sim \alpha_\pi^+ =1.1$ and that at the same momentum $\alpha_p = 1.3$ for protons \cite{adler}.

In Figs. 10 - 13 we have simply multiplied the  $pp$   parameterizations of Badhwar and Mokhov  by a constant  Cronin enhancement factor, $12^\alpha$,
 and compared them to the $pC$ data from NA49 \cite{altpc}.  In our previous $pp$ fits, the Badhwar parameterization worked best for high $p_T > 0.5$ GeV and the Mokhov parameterization worked best for low  $p_T  <  0.5$ GeV.  Therefore we do {\em not} expect the Badhwar parameterization to fit low $p_T$ data for the $pC$ reaction and we do not expect the Mokhov parameterization to  fit high $p_T$ data for the $pC$ reaction.
 We varied the value of $\alpha$ to find the best possible fit to the $pC$ data.
 Figs. 10 - 11 show the Badhwar fit  to the high $p_T$ data with a best value of $\alpha = 0.9$ and   Figs. 12 - 13 show the Mokhov  fit   to the low  $p_T$ data also with an independent  best value of $\alpha = 0.9$.

\section{CONCLUSIONS}

The NA49 collaboration \cite{altpp, altpc} has provided new precise data on pion production in $pp$ and $pC$ reactions at a beam momentum of 158 GeV. Although a precise fit is not expected, nevertheless it is of interest to compare currently available arithmetic parameterizations to the new data. 
Let us emphasize that we are not suggesting that arithmetic parameterizations are able to give a complete account of the new data  {\em et al} \cite{altpp, altpc}. The numerical interpolation developed by the NA49 collaboration \cite{altpp, altpc}  is far superior. The aim of this paper has rather been to see how well some  parameterizations describe the data.

We conclude that the  Alper and Ellis
parameterizations do not fit the $pp$ data very well, and should not be used in this energy region. The Carey parameterization  for $\pi^-$ works better but underpredicts the data at low $p_T$ for small values of $x_F$ and the predictions at high $p_T$ are only moderately good. The Badhwar parameterization for $\pi^{\pm}$  works well for high  values of $p_T$, but does a poor job at low $p_T$, whereas the Mokhov parameterization  is the other way around. It works best for  low
$p_T$ but not well for high $p_T$. Note also that the Badhwar parameterization for $\pi^-$ does not work well for large values of $x_F$, whereas the Mokhov parameterization  works fine in this region.
Regarding the $pp$ parameterizations, we conclude that for low  $p_T < 0.5$ GeV, it is best to use the Mokhov parameterization. For high  $p_T  >  0.5$ GeV,  is best to use the Badhwar parameterization, except for  large values of $x_F >0.45$, where it is better to use  Mokhov.

The Badhwar and Mokhov parameterizations gave good fits in certain $p_T$ ranges and  we scaled them to the $pC$ data.  In the Cronin effect,  the scaling factor is $A^{\alpha(p_T)}$.  We found that the Badhwar parameterization gave the best fit for    $\alpha \approx 0.9$, and  the Mokhov parameterization also gave the best fit for  $\alpha \approx 0.9$. As discussed above, the Badhwar parameterization works best for high $p_T$ and the Mokhov  parameterization works best for low $p_T$.  Therefore we conclude that the Cronin enhancement factor for the $pC$ reactions is 
$\alpha(p_T) \approx 0.9$  in both the high  $p_T  >  0.5$ GeV region and 
 low  $p_T  < 0.5$ GeV region. These results indicate the absence of nuclear collective effects \cite{adler} for this  $pC$ reaction \cite{altpc}.

\noindent 
{\bf Acknowledgments}.
This work was supported by
NASA grant NAG8-1901.

\end{document}